\documentclass[twocolumn,,
 reprint,
 superscriptaddress,
 amsmath,amssymb,
 aps
]{revtex4-1}

\usepackage[dvipdfmx]{graphicx}% 
\usepackage{dcolumn}% 
\usepackage{bm}% 
\usepackage{color}%
\usepackage{siunitx}
\usepackage{here}
\usepackage{mathtools}

\newcommand*\patchAmsMathEnvironmentForLineno[1]{
  \expandafter\let\csname old#1\expandafter\endcsname\csname #1\endcsname
  \expandafter\let\csname oldend#1\expandafter\endcsname\csname end#1\endcsname
  \renewenvironment{#1}
     {\linenomath\csname old#1\endcsname}
     {\csname oldend#1\endcsname\endlinenomath}}
\newcommand*\patchBothAmsMathEnvironmentsForLineno[1]{
  \patchAmsMathEnvironmentForLineno{#1}
  \patchAmsMathEnvironmentForLineno{#1*}}
\AtBeginDocument{
\patchBothAmsMathEnvironmentsForLineno{equation}
\patchBothAmsMathEnvironmentsForLineno{align}
\patchBothAmsMathEnvironmentsForLineno{flalign}
\patchBothAmsMathEnvironmentsForLineno{alignat}
\patchBothAmsMathEnvironmentsForLineno{gather}
\patchBothAmsMathEnvironmentsForLineno{multline}
}
\begin{document}
\preprint{APS/123-QED}

\title{Reverse heat flow with Peltier-induced thermoinductive effect}% 

\author{K. Okawa}
%\email{okawa.k@aist.go.jp}
\affiliation{National Institute of Advanced Industrial Science and Technology (AIST), National Metrology Institute of Japan (NMIJ), Tsukuba, Ibaraki, Japan 305-8563}
\author{Y. Amagai}%
\affiliation{National Institute of Advanced Industrial Science and Technology (AIST), National Metrology Institute of Japan (NMIJ), Tsukuba, Ibaraki, Japan 305-8563}
\author{H. Fujiki}%
\affiliation{National Institute of Advanced Industrial Science and Technology (AIST), National Metrology Institute of Japan (NMIJ), Tsukuba, Ibaraki, Japan 305-8563}
\author{N-H. Kaneko}%
\affiliation{National Institute of Advanced Industrial Science and Technology (AIST), National Metrology Institute of Japan (NMIJ), Tsukuba, Ibaraki, Japan 305-8563}

%\date{\today}

%\begin{abstract}
%\end{abstract}

\maketitle
%\tableofcontents
{\bf The concept of “thermal inductance” expands the options of thermal circuits design. However, the inductive component is the only missing components in thermal circuits unlike their electromagnetic counterparts. Herein, we report an electrically controllable reverse heat flow, in which heat flows from a low-temperature side to a high-temperature side locally and temporarily in a single material by imposing thermal inertia and ac current. This effect can be regarded as an equivalent of the “thermoinductive” effect induced by the Peltier effect. We derive the exact solution indicating that this reverse heat flow occurs universally in solid-state systems, and that it is considerably enhanced by thermoelectric properties. A local cooling of $\SI{25}{mK}$ is demonstrated in (Bi,Sb)$_{2}$Te$_{3}$, which is explained by our exact solution. This effect can be directly applicable to the potential fabrication of “thermoinductor” in thermal circuits.}

\begin{figure}[t]
\centering
\includegraphics[width=8cm,clip]{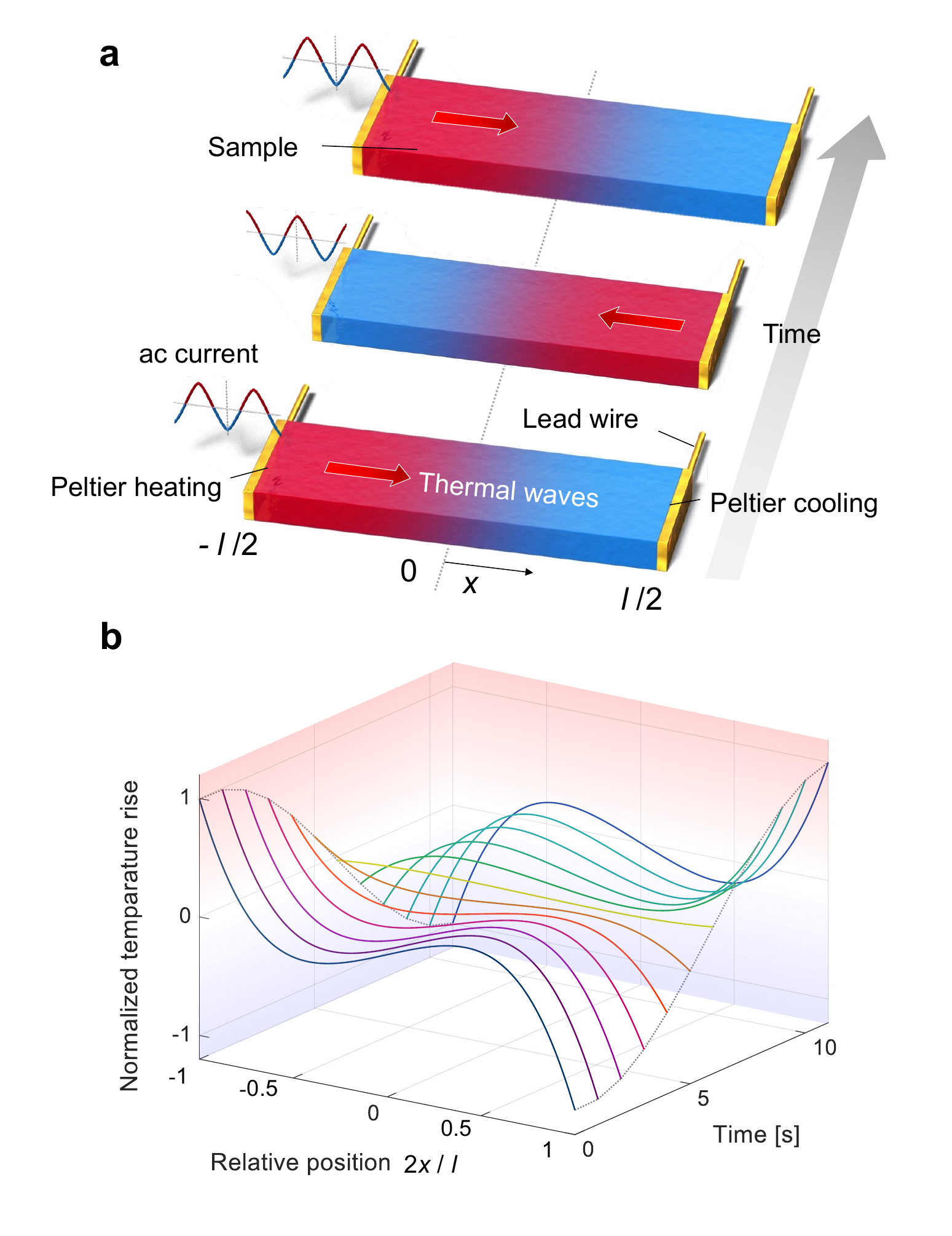}
\caption{{\bf Temperature change caused by the Peltier effect with periodic modulation.}  {\bf a} Schematic of changes in temperature of the material. By applying an ac current, Peltier heating and cooling are generated at the edge of the sample and the thermal oscillation propagates in the material as a thermal wave. The direction of the wave is changed by reversing the polarity of the ac current. $l$ and $x$ are the length of the sample and the position in the sample, respectively. {\bf b} Calculation results of the temperature rise caused by the Peltier effect with periodic modulation at the edge of the sample as a function of relative position in the sample $2x/l$ and time in the case of the characteristic dimensionless parameter $F = 1$. Here, $F$ is the product of the thermal time constant of the sample and the current frequency $(F \equiv l/2)2f/2\alpha$, where $f$ and $\alpha$ are the current frequency and the thermal diffusivity of the sample,
respectively. The vertical axis is normalized by the value at the edge, $2x/l = -1$.}
\label{fig1}
\end{figure}
Thermal design optimization is an essential issue that needs to be addressed for the development of high-performance electronic devices\cite{Garimella1,Ma2,Ahmed3,Shen4}. To date, for the realization of a versatile platform for thermal control in solid-state devices, novel thermal components that can manipulate phonon transport (i.e., thermal diodes\cite{Li5,Ben6,Martinez7}, transistors\cite{Li8,Sood9}, logic gates\cite{Wang10}, and memories\cite{Wang11,Kubytskyi12,Guarcello13}) have been proposed and discussed\cite{Li14,Wehmeyer15,Ye16}. These thermal concepts are motivated by the idea of familiar electronic analogs.

An electrical current flow can be essentially replaced with a heat flow by analogy with the Ohm’s law\cite{Wehmeyer15,Bosworth17,Weedy18,Lienhard19,Piggott20}; however, an “inductor” is missing among components for thermal circuits\cite{Wehmeyer15,Lienhard19}. This is because oscillatory behaviors, with a reversal in the direction of heat flow from cold to hot, are typically considered violations of the second law of thermodynamics\cite{Weedy18,Bossen21}. An equivalent of the “thermoinductive” effect has been reported in circuits provided with unclear and cumbersome interventions using external heat flows (i.e., natural convection)\cite{Bosworth22,Bosworth23}, or the cooling of an electrical coil to the temperature of liquid He\cite{Schilling24}.

Herein, we report an electrically controllable reverse heat flow induced by the Peltier effect with periodic modulation, which operates according to an external ac current, generated locally and temporarily in a single thermoelectric (TE) material. The underlying concept is the heating and cooling of the ends of the material by the Peltier effect under an applied ac current; this forms a negative temperature gradient locally and temporarily in the opposite direction in a controllable manner. The high controllability of heat flow via manipulating the electrical current has promising potential applications. The circuit, composed of a single material, facilitates the understanding of reverse heat flow through theoretical modeling using a simple heat conduction model. Based on the exact solution derived from our heat transfer analysis, we determined an optimized condition to enhance the reverse heat flow occurring in a material. To show proof of concept, we measured and detected the electrical resistance change caused by the reverse heat flow, which is compelling evidence for the Peltier-induced thermoinductive effect, achieved using an excellent TE material, (Bi,Sb)$_{2}$Te$_{3}$, near room temperature.\\

\begin{figure*}[t]
\centering
\includegraphics[width=12.5cm,clip]{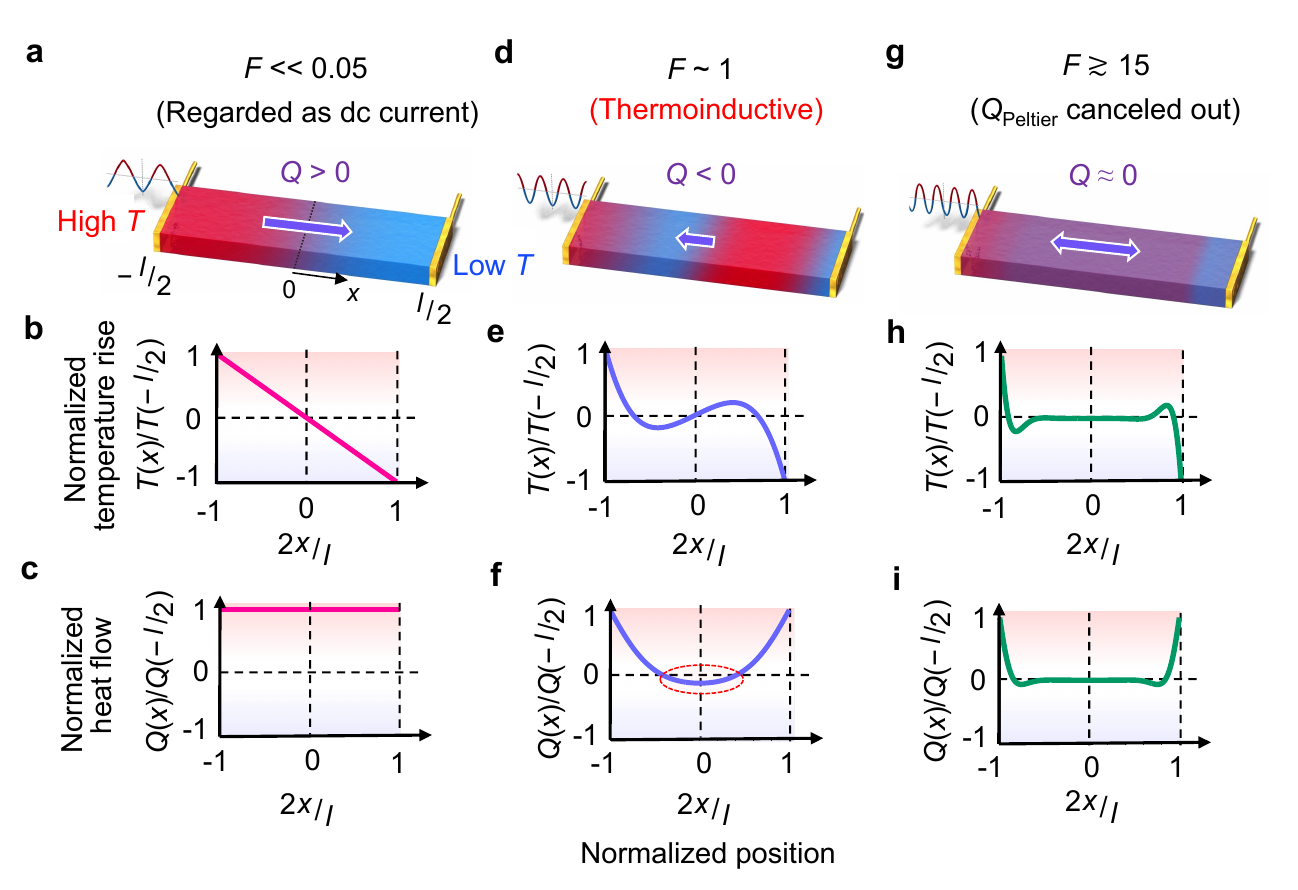}
\caption{{\bf Classification of the temperature distribution and heat flow induced by Peltier effect.} Schematic of the temperature distribution $T(x)$ and heat flow $Q(x)$ in the sample at the time under the applied rms ac current. The situation is classified by the characteristic dimensionless parameter $F$ representing the product of the thermal time constant of the sample and the current frequency ($F \equiv (l/2)2f/2\alpha$, where $f$ and $\alpha$ are the current frequency and the thermal diffusivity of the sample, respectively), as (a)--(c) $F \ll 0.05$, (d)--(f) $\sim 1$, and (g)--(i) $\gtrsim 15$. (a), (d), (g) The schematic of the temperature distribution corresponding to each $F$ region in the material are shown. The relationship (b), (e), (h) between $T(x)$ and relative position in the sample $2x/l$, and (c), (f), (i) between $Q(x)$ and $2x/l$ are also shown, here $l$ is the length of the sample. In the case of $F \sim 1$, the thermoinductive effect (reverse heat flow) occurs because of the Peltier effect.}
\label{fig2}
\end{figure*}

\par
\noindent {\bf  Results}\\
{\bf Modeling of the Peltier-induced thermoinductive effect.} In the rectangular parallelepiped TE material shown in Fig.~1a, Peltier heating and cooling occur at each interface between the material and the metal electrode under an applied current. Consequently, the direction of the heat flow transmitted as a thermal wave is also reversed with the sign change of the ac current \cite{Iwasaki25,Downey26}. The reversing of the current polarity causes thermal wave collision depending on the position in the material (Fig.~1b). An opposite temperature gradient (i.e., a reverse heat flow) can occur in the material at a certain frequency at which the time for reversing the polarity of
the ac current is sufficiently faster than the thermal time constant of the material (see the Supplementary Movie 1). We derived an exact solution for the heat conduction equation to reveal the temperature distribution in the material.

The temperature rise caused by the Peltier effect $T(x,t)$ at position $x$ and time $t$ is described as
\begin{eqnarray}
T(x,t) = \dfrac{SJT_{0}}{\kappa} \dfrac{1-i}{2\beta} \dfrac{\mathrm{e}^{(1+i)\beta x}-\mathrm{e}^{-(1+i)\beta x}}{\mathrm{e}^{(1+i)\beta l/2}+\mathrm{e}^{-(1+i)\beta l/2}},
\end{eqnarray}
where $S$, $T_{0}$, $\kappa$, and $l$ are the Seebeck coefficient, mean temperature, thermal conductivity, and length of the sample, respectively. $J$ is the current density given by $J = J_{0}\sin{\omega t}$ ($\omega$ is the angular velocity $2\pi f$; $f$ is the current frequency).  $\beta \equiv  \sqrt{\omega/2\alpha}$ is the reciprocal of the thermal diffusion length $D_{th} = \sqrt{\alpha/\pi f}$, where $\alpha$ is the thermal diffusivity of the sample. $i$ is the imaginary number. The imaginary part corresponds to the phase of the thermal wave. Equation~(1) is the exact solution for the one-dimensional unsteady-state heat transfer equation (see Methods for calculation details).

Here, we introduce a characteristic dimensionless parameter $F \equiv (l/2)^{2}f /2\alpha$, representing the product of the thermal time constant of the sample $(l/2)^{2}/\alpha$ and the current frequency. We proposed that $F$ can classify the behavior of the temperature distribution caused by the Peltier heat. Figure~2 shows the relationship between $T(x)$ or heat flow  $Q(x)$ = $-\kappa$$A$d$T$/d$x$, where $A$ is the cross-sectional area, and the normalized position $2x/l$. $T(x)$ and $Q(x)$ represent the values at the time under the applied root-mean-square (rms) ac current. $T(x)$ and $Q(x)$ can be classified into three types, the dc current region (Figs.~2a--c), the thermoinductive region (Figs.~2d--f), the region where Peltier heat cancels out (Figs.~2g--i), respectively. When the current can be regarded as almost dc current (i.e., $F~\ll~0.05$), a linear temperature gradient is caused by the Peltier heating and cooling generated at each sample edge. When the ac current of a sufficiently high frequency cancels the TE effect (i.e., $F~\gtrsim~15$), the temperature distribution is flat, except at the sample edges. In contrast, a temperature difference with the opposite direction in the sample occurs when $F \sim 1$. The direction of the temperature gradient is periodically inverted with the polarity reversal of the current (Fig. 1b). As can be seen in Fig. 2h, $Q$ with the opposite (negative) direction occurs near the sample center ($x \sim 0$). This uneven heat flow occurring within the material represents a thermal phase delay against the current. Although it is a local and temporary effect, this reverse heat flow can be interpreted as thermoinductive effect induced by the Peltier effect. 

In addition, our calculation based on the exact solution, Eq.~(1) shows that while a reverse heat flow can universally occur in any solid material, it is more prominent in TE materials (Fig.~3a,~b). Here, the typical physical properties of (Bi,Sb)$_{2}$Te$_{3}$ ($\alpha = \SI{1.25e-6}{m^{2}s^{-1}}$, $S=\SI{165}{\micro VK^{-1}}$, and $\kappa = \SI{1.45}{Wm^{-1}K^{-1}}$) and Cu ($\alpha = \SI{1.17e-4}{m^{2}s^{-1}}$, $S = \SI{6.5}{\micro VK^{-1}}$, and $\kappa~=~\SI{398}{Wm^{-1}K^{-1}}$) at room temperature were used in the calculation. The negative temperature gradient occurs in both materials, showing a temperature decrease reaching $\sim \SI{25}{mK}$ at rms ac current $I_{\rm rms}~=~\SI{10}{mA}$ and $\SI{44.4}{mHz}$ in (Bi,Sb)$_{2}$Te$_{3}$. A partial temperature inversion of $\sim \SI{20}{\%}$ of the Peltier heating can be achieved. Conventional metallic Cu shows a temperature decrease of only $\SI{3.5}{\micro K}$ at $I_{\rm rms}~=~\SI{10}{mA}$ and $\SI{4.16}{Hz}$, which is ten thousand times smaller than that shown by (Bi,Sb)$_{2}$Te$_{3}$. The large Seebeck coefficient is key to enhancing the reverse heat flow. The low thermal conductivity also allows control of the reverse heat flow at low frequencies (Supplementary Fig.~1 and Supplementary Table~1). However, the direct and high-accuracy detection of minute changes in the heat flow (on the order of $\SI{}{\micro W}$) is difficult hindering verification of this phenomenon using general calorimetric measurements.\\

\noindent{\bf Electrical resistance changes caused by thermoinductive effect.}
We present the analytical model for electrical impedance measurements with a four-probe configuration that can detect temperature changes reflected in the electrical voltage by the Seebeck effect (see Methods). The temperature changes caused by the reverse heat flow can be measured as electrical signals on the order of m$\Omega$ using TE materials. The measured impedance $R$ can be expressed as follows with the ohmic resistance $R_{0} = \rho l_{\rm V}/A$:
 \begin{eqnarray}
R = R_{0}[1+zT_{0}(R_{1}+iR_{2})],
\end{eqnarray}
where $\rho$ and $l_{\rm V}$ denote the resistivity and voltage terminal distance. $z$ is the TE figure of merit where $z = S^{2}/\kappa \rho$. Here, $R_{1}$ and $R_{2}$ are the following functions:

\begin{figure}[t]
\centering
\includegraphics[width=7.5cm,clip]{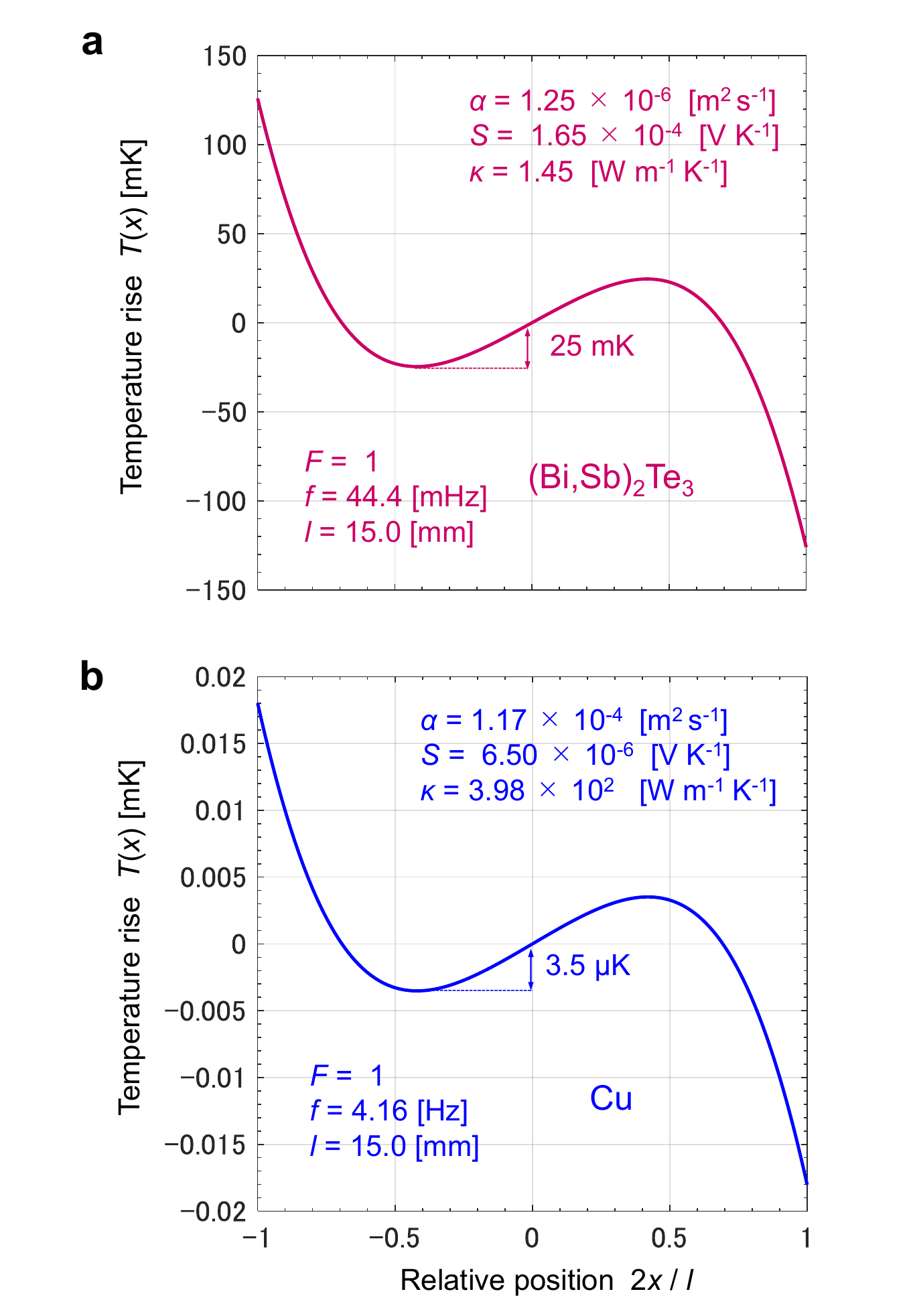}
\caption{{\bf Temperature change caused by the Peltier effect with periodic modulation.} Calculation results based on the exact solution of the temperature rise $T(x)$ dependence of relative position in the sample $2x/l$ at the time under the applied root-mean-square (rms) ac current in {\bf a} a thermoelectric material (Bi,Sb)$_{2}$Te$_{3}$ and {\bf b} conventional metallic Cu for the characteristic dimensionless parameter $F = 1$. Here, $F$ is the product of the thermal time constant of the sample and the current frequency ($F \equiv (l/2)2f/2\alpha$, where $f$ and $\alpha$ are the current frequency and the thermal diffusivity of the sample, respectively). Typical physical properties of (Bi,Sb)$_{2}$Te$_{3}$ ($\alpha = \SI{1.25e-6}{m^{2} s^{-1}}$, $S = \SI{165}{\micro VK^{-1}}$, and $\kappa = \SI{1.45}{Wm^{-1}K^{-1}}$) and Cu ($\alpha = \SI{1.17e-4}{m^{2}s^{-1}}$, $S = \SI{6.5}{\micro V K^{-1}}$, and $\kappa = \SI{398}{Wm^{-1}K^{-1}}$) at room temperature were used in the calculation, where $S$ and $\kappa$ is the Seebeck coefficient and thermal conductivity of the sample, respectively.}
\label{fig3}
\end{figure}

\begin{widetext}
\begin{eqnarray}
R_{1,2} = \dfrac{\cos{\mu}\cosh{\mu}(\sin{\nu}\cosh{\nu}\pm \cos{\nu}\sinh{\nu})\pm \sin{\mu}\sinh{\mu}(\sin{\nu}\cosh{\nu}\mp \cos{\nu}\sinh{\nu})}{2\nu[(\cos{\mu}\cosh{\mu})^{2}+(\sin{\mu}\sinh{\mu})^{2}]} 
\end{eqnarray}
\end{widetext}
where $\mu~\equiv~\sqrt{2\pi F}$ and $\nu~\equiv~\mu l_{\rm V}/l$ are defined as functions depending on $F$ and $l_{\rm V}/l$. The second term on the right side in Eq.~(2) represents the component resulting from the TE effect, which is increased or decreased by the correction terms $R_{1}$ and $R_{2}$. $R_{1}$ and $R_{2}$ are considered to correspond with the resistance and reactance components of the TE voltage, respectively. Therefore, $R_{1} \rightarrow 1$ and $R_{2} \rightarrow 0$ when $F \rightarrow 0$ as a dc limit, and $R_{1} \rightarrow 0$ and $R_{2} \rightarrow 0$ when $F \rightarrow \infty$ when an ac current with a sufficiently high frequency is applied. The influence of $R_{1}$ and $R_{2}$ can be better observed using high-performance TE materials. Figure~4a,~b shows the calculation results regarding the dependency of the correction terms $R_{1}$ and $R_{2}$ on $F$ obtained using the exact solution of Eq.~(3). The inset of Fig.~4b shows the Nyquist plot for $R_{1}$ and $R_{2}$. Here, the physical properties of (Bi,Sb)$_{2}$Te$_{3}$ at room temperature were used. At low current frequencies with $F < 10^{-2}$, $R_{1}$ approaches 1 and the TE effect is sufficiently generated in the sample, but it gradually decreases to 0 as $F$ increases. Notably, as the voltage probe position normalized by the sample length $l_{\rm V}/l$ becomes smaller, a large dip structure of $R_{1}$ occurs $F \sim 1$. This reflects the situation in which the temperature distribution due to the reverse heat flow shown in Fig.~2e is generated in the sample. The dip structure is not measured in the two-probe measurement as $l_{\rm V}/l = 1$. This is consistent with the absence of the dip structure in reports of electrical impedance measurement with the two-probe configuration,
such as a $zT$ estimation using an impedance spectroscopy technique \cite{Garcia27,Beltrani28,Shinozaki29}. Our exact solution revealed that the size of the reverse heat flow varied depending on $x$, and that it could be detected through electrical impedance measurements with changing $l_{\rm V}$.

\begin{figure*}[t]
\centering
\includegraphics[width=13cm,clip]{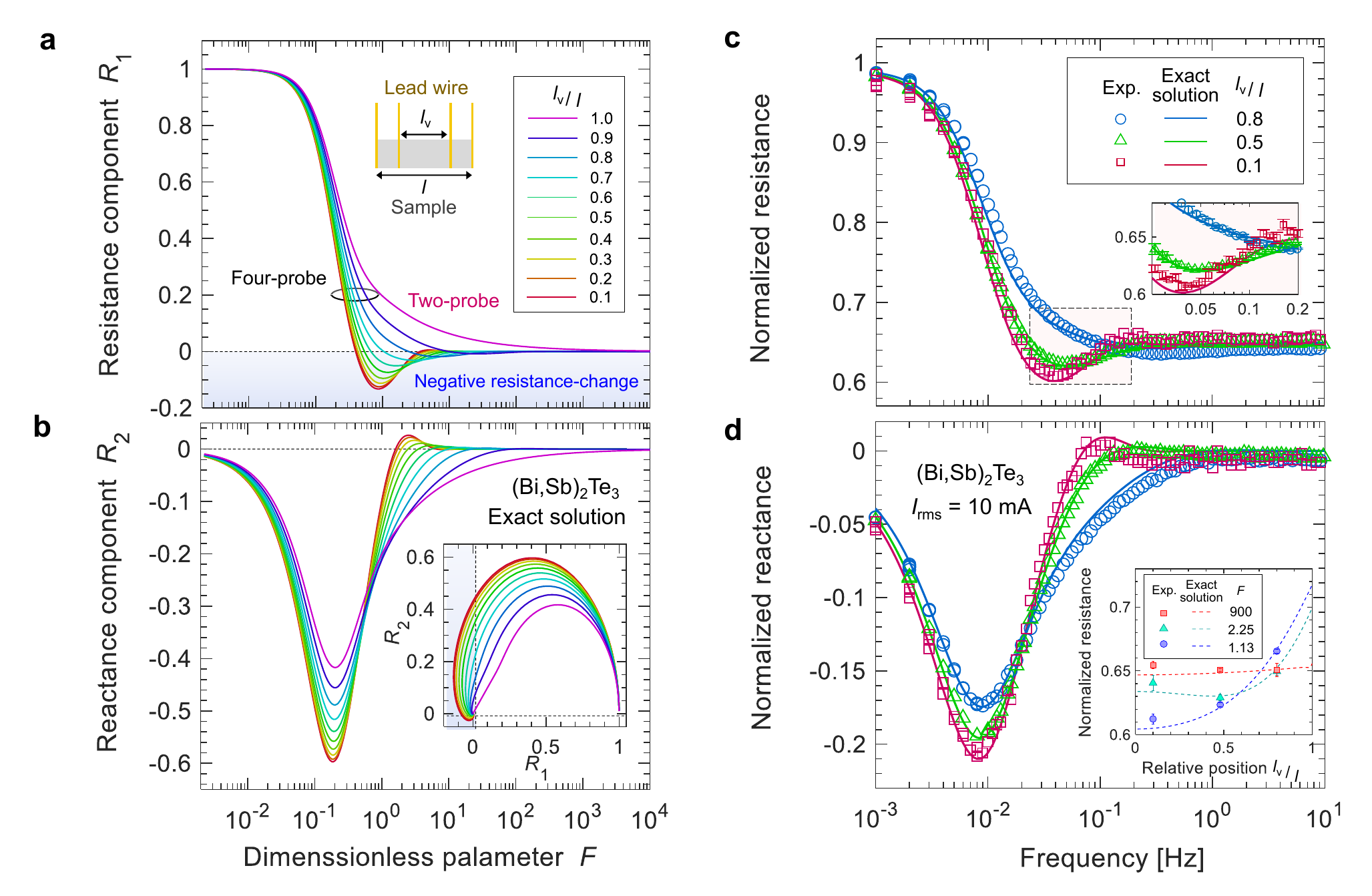}
\caption{{\bf Calculation and experimental results of Peltier-induced thermoinductive effect.} Calculation results based on the exact solution regarding the dependence of characteristic dimensionless parameter $F$ representing the product of the thermal time constant of the sample and the current frequency ($F \equiv (l/2)^{2}f/2\alpha$, where $l$, $f$, and $\alpha$ are the sample length, the current frequency and the thermal diffusivity of the sample, respectively) on the correction terms {\bf a} $R_{1}$ and {\bf b} $R_{2}$ using (Bi,Sb)$_{2}$Te$_{3}$ with $\alpha = \SI{1.25e-6}{m^{2}s^{-1}}$, the Seebeck coefficient $S = \SI{166}{\micro VK^{-1}}$, the electrical resistivity $\rho = \SI{1.1e-5}{\Omega m}$, the thermal conductivity $\kappa = \SI{1.45}{Wm^{-1}K^{-1}}$, and $l = \SI{15}{mm}$. $R_{1}$ and $R_{2}$ are considered to correspond with the resistance and reactance components of the thermoelectric voltage. The voltage probe position, $l_{\rm V}$, normalized by the sample length $l_{\rm V}/l$ was varied from 0.1 to 1.0. Inset in a shows the illustration of the four-probe configuration. Inset in b shows the Nyquist plot using the exact solution. The blue region denotes the negative resistance change that has occurred. Experimental results of frequency dependence of the {\bf c} real and {\bf d} imaginary parts of the impedance of (Bi,Sb)$_{2}$Te$_{3}$. Open symbols are experimental results, and solid lines are calculation results based on the exact solution, Eq.~(2). Red circles, green triangles, and purple squares denote the experimental results corresponding to $l_{\rm V}/l = 0.8$, 0.5, and 0.1, respectively. The vertical axis is normalized by the value for the dc measurement ($F \rightarrow 0$). The inset in {\bf c} shows the enlarged view on the plot with error bars in the negative resistance-change region. The error bars show the standard deviation of the mean for repeated measurements of the resistance. The inset in {\bf d} shows the relationship between the normalized resistance and $l_{\rm V}/l$. Dotted lines are calculation results. Red squares, green triangles, and blue circles are the experimental results corresponding to $F = 900$, 2.25, and 1.13, respectively.}
\label{fig4}
\end{figure*}

To verify the validity of the analytical model, electrical impedance measurements using the TE material (Bi,Sb)$_{2}$Te$_{3}$ were conducted with a four-terminal pair configuration (see Methods for details of the experimental setup). The measurements were performed by changing $l_{\rm V}/l$ from 0.8 to 0.1. Figure~4c, d shows the frequency dependence of the real and imaginary parts of the measured electrical impedance. The resistance values were gradually decreased with an increase in the current frequency at $l_{\rm V}/l = 0.8$. When $l_{\rm V}/l  = 0.5$ and 0.1, a notable dip structure was
observed around $\SI{40}{mHz}$. This frequency region roughly corresponds to $F = 1$ in this material. The resistance value for a flat region, corresponding to $R_{0}$, was obtained because the TE effect was sufficiently canceled at frequencies exceeding $\SI{500}{mHz}$. As expected, the calculation results indicated that the size of the dip structure increased with a decreasing $l_{\rm V}/l$. The plots developed using the exact solution, Eq.~(2), agreed well with the experimental results over a wide frequency range. The inset of Fig.~4c shows the enlarged view around $\SI{50}{mHz}$ with error bars. The error bars show the standard deviation of the measurement results of the electrical impedance with a four-pair terminal configuration. In the region where negative resistance changes occur, the difference between the measured impedance values with each $l_{\rm V}/l$ can be clearly identified even with error bars. The inset of Fig.~4d shows the relationship between the normalized resistance and $l_{\rm V}/l$. At $F = 900$ ($f=\SI{40}{Hz}$), which is in the flat frequency region in Fig.~4c, the value of the normalized resistance does not change; this is because of the normalized $R_{0}$, which is $\sim 0.65$. The maximum difference from $R_{0}$ is $\sim 6.5 \%$ at $l_{\rm V}/l = 0.1$ in (Bi,Sb)$_{2}$Te$_{3}$. Conventional metallic Cu requires detecting minute changes on the order of n$\Omega$ or less thorough electrical resistivity measurements, which is a challenging task (see Supplementary Fig.~1).\\

\noindent {\bf  Disscussion}\\ To investigate the conditions under which the reverse heat flow appears, we discussed the heat flow behavior in the frequency domain. In this study, a reverse heat flow is generated by utilizing the material’s thermal inertia and the phase delay caused by the external current reversal with periodic modulation. Here, “thermal inertia” is a thermal response of a material, generally pertaining to the product of thermal resistance and volume-specific heat capacity, and causes a delay in the rate of heat flow through the material. In the present case, the delay of heat flow caused by the thermal inertia by itself does not produce the reverse heat flow, but the reversal Peltier heating and cooling induced by the current reversal causes the negative heat flow in the center of the material. Therefore, the behavior of $Q(F)$ can be separable into the resistor-capacitor (RC) component that originates from the
thermal inertia and the oscillation component that is induced by the ac current (Fig.~5). The denominator of the expression of $Q$ (see Eq.~(9) in the Method section) corresponds to the RC component as a component independent of $x$ and sharply decreases at $F > 0.1$ because of the thermal inertia (the blue dashed line in Fig.~5). In the case of two-probe measurements ($l_{\rm V}/l = 1$), the behavior of $Q(F)$ can be explained by considering only this RC component. The component obtained by subtracting the RC component from $Q(F)$ is responsible for the oscillation of $Q$ with respect to $F$ (the green dashed line in Fig.~5). This oscillation component substantially contributes to $Q(F)$ for $F > 0.1$, causing a significant reverse heat flow at $F \sim 1$.

In addition, even after $Q(F)$ becomes negative once, $Q(F)$ oscillates with increasing $F$ while repeating its sign inversion. Because the value of $F$ at which the reverse heat flow appears corresponded to that in the case where $Q(F) = 0$, the first negative $Q$ appeared in the region of $\pi/8 < F < 9\pi/8$. Thereafter, the sign of $Q$ inverts in the $\pi$ cycle, which corresponds to the cycle of the current's reversing polarity. Further, we find that when $\mu = (l/2)/ D_{\rm th} = (2n-1)\pi/2, (n = 1, 2,...)$, the sign of $Q$ is inverted. Therefore, we infer that the reverse heat flow in materials can be controlled by the ratio of the sample length and $D_{\rm th}$. This thermoinductive effect is a higher-order thermal response due to the Peltier effect, which cannot be obtained in a conventional lumped-parameter model. 

A Peltier-induced reverse heat flow has the advantage of easy tunability using an external current source, in contrast with previously proposed circuits, using external electrical coils \cite{Bossen21,Schilling24} and natural convection \cite{Bosworth22,Bosworth23}. For example, the control of heat flow in a fluid system is difficult because the exact analytical model and required experimental conditions are still unclear. Alternatively, the electrical coil with a suitable inductance and internal resistance can be used to adjust the heat flow. In such a case, incorporation of a superconducting coil operating at liquid He temperatures in the circuit is desirable to reduce the internal resistance of the electrical coil. This impairs the practical implementation of such approaches in thermal circuits. Arbitrary reverse heat flow can be generated without replacing circuit components by tuning the current frequency regardless of the sample size and physical properties (i.e., thermal diffusivity). 

We note that a transient underdamped oscillation of heat flow is realized excellently by integrating an electrical coil (a superconducting coil) into the circuit in the previous report \cite{Schilling24}. The “thermal inductor” shown in the report is composed of the thermoelectric module and the electric inductor. In analogy to the self-inductance $L$ of an electrical coil, they describe the “thermal self-inductance” using a lamped-parameters model, $L_{\rm th} \sim L/(S^{2}T$), where $S$ and $T$ are the Seebeck coefficient of the device and the temperature at one end of the device. This formula can be shown to be analogous to the electric LCR circuit in their system. However, this is derived from the thermal balance equation in contrast to our exact solution derived from the heat transfer equation. In our case, the “local” and “temporal” reverse heat flow is dealt with a time-dependent heat transfer equation. Describing the reverse heat flow in terms of the steady-state lumped-parameter model and defining clearly the $L_{\rm th}$ will be very difficult. If such a meaningful thermal self-inductance is defined, the novel thermal circuit design such as a “thermal resonant circuit” would be realized, which will greatly expand the potential of the thermal design. 

\begin{figure}[t]
\centering
\includegraphics[width=8cm,clip]{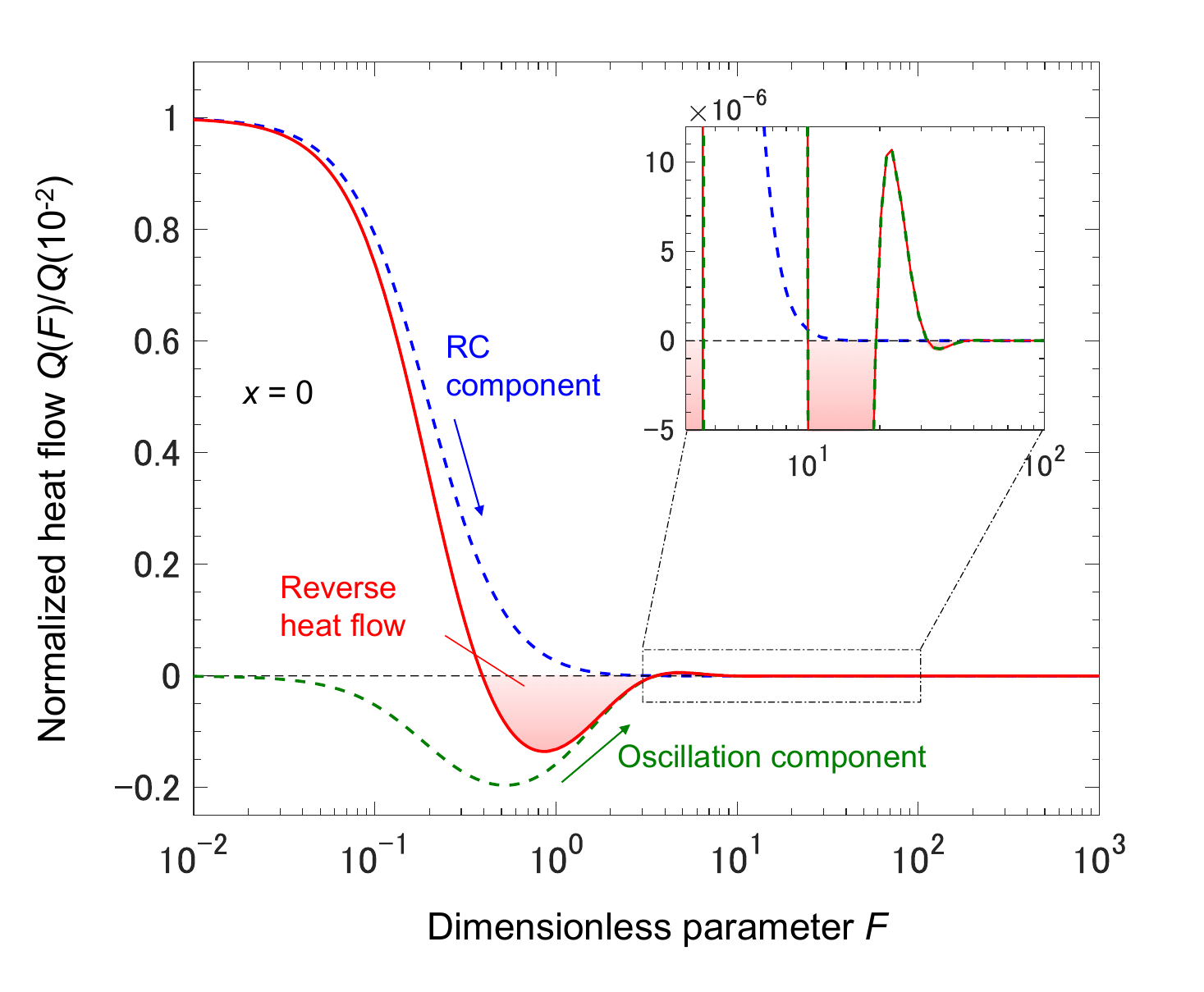}
\caption{{\bf Characteristic dimensionless parameter $F$ region in which reverse heat flow ($Q < 0$) is expected.} alculation results showing the dependence of the characteristic dimensionless parameter $F$, representing the product of the thermal time constant of the sample and the current frequency ($F \equiv (l/2)2f/2\alpha$, where $f$ and $\alpha$ are the current frequency and the thermal diffusivity of the sample, respectively), on the heat flow $Q(F)$ at $x = 0$. The vertical axis is normalized by the value at $F = 10^{-2}$. The solid red line is the heat flow $Q(F)$ calculated using Eq.~(9). The blue dotted line is the attenuation term of $Q(F)$, corresponding to the resistor-capacitor (RC) component towing to the material’s thermal inertia. The green dotted line is the oscillation term of $Q(F)$, corresponding to the oscillation component induced by the external current reversal. The inset shows an enlarged view of the damped oscillation in the larger $F$ region.}
\label{fig5}
\end{figure}

In conclusion, we succeeded in analytically formulating and experimentally observing a Peltier-induced reverse heat flow, which can be regarded as an equivalent to the thermoinductive effect, in a single TE material. The exact solution obtained by heat transfer analysis clearly demonstrated the optimal conditions for realizing and enhancing this reverse heat flow, induced by the thermal inertia and the phase delay under an external ac current. The excellent agreement between the exact solution and the experiments indicated that the reverse heat flow could be well controlled by the utilization of heat flow with the variable frequency current. A partial temperature inversion of $\sim 20~\%$ of the Peltier heating was achieved temporarily at room temperature. These results demonstrate that the utilization of TE materials is a promising strategy for manipulating heat flow in thermal circuits. Our study contributes toward understanding the mechanisms of equivalents of the thermoinductive effect and thereby broadens the scope for thermal circuit design in electronic devices.

\vspace{5mm}
\noindent {\bf  Methods}\\
{\bf Experimental setup}. Impedance was measured using a polycrystalline p-type (Bi,Sb)$_{2}$Te$_{3}$ material. The dimensions of the sample were $\SI{15}{mm} \times \SI{4}{mm} \times \SI{1}{mm}$ as shown in Supplementary Fig.~2. To reduce the thermal interface resistance between the sample and electrode, $\SI{0.2}{\micro m}$ thick Au electrodes were fabricated by a magnetron sputtering. Upon confirming that the adhesion between the sample and the lead wires was insufficient, the measured dc voltage was found to have a large deviation. To suppress the heat conduction of the lead wire, thin Au wires ($\Phi = \SI{30}{\micro m}$) were connected to the sample using Ag paste. The Ag paste was cured sufficiently (20~minutes at $\SI{298}{K}$) according to the specifications. The sample was then quickly placed into the sample space under vacuum conditions. This process was done in the same way for each measurement. After curing, no significant change in the electrical resistance was observed, even when the sample was removed from the sample space and re-measured. The need of a homogeneous Peltier effect at the junctions and the effect of the contact resistance arising from the use of Ag paste can be error sources. This problem has been pointed out in the previous reports using the Harman method and the impedance spectroscopy method, which are measurement techniques for evaluating thermoelectric properties of the solid-state materials with two- or four-probe configurations \cite{Beltrani28}. To isolate the sample thermally, the sample was suspended with sufficiently long Au wires. The heat loss of thermal radiation is not negligibly small on measurements dealing with thermoelectric effects, even at room temperature, depending on the geometry of the sample \cite{Amagai31}. The sample space was covered with a circular radiation shield made of oxygen-free Cu to reduce heat loss caused by thermal radiation (see Supplementary Fig.~2). The measurement apparatus was assembled in a vacuum chamber, and measurement was performed under high vacuum conditions ($\SI{e-3}{Pa}$ or less) to suppress the influence of heat convection. The impedance was measured using an impedance analyzer (HIOKI, IM3590). Coaxial cables were used to connect the case, and measurements were performed using the four-terminal pair method. A schematic of the measurement setup in this study is shown in Supplementary Fig.~2. Although we had to ensure that the Joule heat did not exceed the Peltier heat, the applied current dependency was measured in advance, and the measurement was performed with a rms ac current of $\SI{10}{mA}$, which did not affect the results in this study (see Supplementary Fig.~3). The impedance analyzer used
in the experiment was calibrated by four-terminal pair standard resistors (Keysight, 42030 A).
\\

\noindent{\bf Complete derivation of the exact solution.} 
The analysis model in this study is described below. We considered the case where the electrical impedance measurement is performed on a rectangular parallelepiped sample with a sample length of $l$ and a cross-sectional area $A$ using the four-terminal pair method with a voltage
terminal distance of $l_{\rm V}$. The Seebeck coefficient, resistivity, thermal conductivity, and thermal diffusivity of the material are denoted as $S$, $\rho$, $\kappa$, and $\alpha$, respectively. These physical parameters are considered temperature independent. The temperature distribution at any position $x$ and time $t$ is described as $T(x,t)$. When an ac current with a current density $J = J_{0}\sin{\omega t}$ is applied, $T(x,t)$ can be obtained by solving the following one-dimensional unsteady-state heat transfer equation \cite{Kirby30}:
\begin{eqnarray}
\frac{\partial^2 T(x,t)}{\partial x^2}=\frac{1}{\alpha}\frac{\partial T(x,t)}{\partial t}.
\label{eq4}
\end{eqnarray}

According to the method of separation of variables, the general solution of Eq.~(4) is described as
\begin{eqnarray}
T(x,t) = C_{1}\mathrm{e}^{\pm(1+i)\beta x-i\omega t}+C_{2}\mathrm{e}^{\mp (1+i)\beta x-i \omega t},
\end{eqnarray}
where $C_{1}$ and $C_{2}$ are arbitrary constants. From Eq. (5), it can be understood that $T(x,t)$ behaves as a thermal wave. The imaginary part corresponds to the phase of the thermal wave. For simplicity, we assume that only Peltier heat $Q_{\rm Peltier}$ = $SJT$$_{0}$ occurs at both ends of the sample $x = \pm l/2$, where $T_{0}$ is the mean temperature of the sample. Assuming that all the Peltier heat flows into the sample, the effects of other heat losses, such as those by convection and radiation, are not considered in this model. Then, the boundary condition is described as follows:
\begin{eqnarray}
\left.\frac{\partial T(x,t)}{\partial x}\right|_{x=\pm l/2} = \frac{SJT_{0}}{\kappa}.
\end{eqnarray}

Under this boundary condition, the arbitrary constants of the general solution of Eq. (5) are obtained. Then, $T$($x$,$t$) is given by Eq.(1).

Using $T(x,t)$ obtained from Eq. (1), the voltage measured between $\pm l_{\rm V}/2$ is expressed as follows:
\begin{eqnarray}
V = l_{\rm V} \rho J + \int_{-l/2}^{l/2} S\dfrac{\partial T(x,t)}{\partial x}dx.
\end{eqnarray}
The first term on the right side represents the ohmic voltage between $l_{\rm V}$, and the second term represents the Seebeck voltage. In actual measurements, the temperature rise caused by the Peltier effect is considered small ($\sim \SI{0.1}{K}$), so the integral part can be replaced with $2ST(x = l_{\rm V}/2, t)$. The measured impedance $R$ in Eq.~(2) can be introduced from $R = J/A$.
A heat flow $Q = -\kappa A$d$T$/d$x$ is described as follows:
\begin{eqnarray}
Q = -JST_{0}A \dfrac{\mathrm{e}^{(1+i)\beta x}+\mathrm{e}^{-(1+i)\beta x}}{\mathrm{e}^{(1+i)\beta l/2}+\mathrm{e}^{-(1+i)\beta l/2}}.
\end{eqnarray}
The real part of $Q$ can be rewritten as follows:
\begin{widetext}
\begin{eqnarray}
Re(Q) =-JST_{0}A\dfrac{\cos{\beta x}\cosh{\beta x}\cos{\mu}\cosh{\mu}+\sin{\beta x}\sinh{\beta x}\sin{\mu}\sinh{\mu}}{\cos^2{\mu}\cosh^2{\mu}+\sin^2{\mu}\sinh^2{\mu}}.
\end{eqnarray}
\end{widetext}
The imaginary part of $Q$ corresponds to the phase of the heat flow. 

In the experimental setup used in this study, only small temperature changes of $< \SI{0.1}{K}$ occurred in the TE material. To increase this effect, the current value can be increased by using a TE material with high $zT_{0}$. However, our model ignores the influence of Joule heating; therefore, if the current value is increased, the effect of Joule heating cannot be ignored. In such a case, a more complicated analysis is required.

\vspace{2.5mm}\noindent
{\bf Acknowledgements}  This work was supported by a Grant-in-Aid for Research Activity Start-up (Grant No.17H07399) from Japan Society for Promotion of Science (JSPS).

\vspace{2.5mm}\noindent
{\bf Author Contributions}  K.O. carried out main experiments and the data analyses. Y.A. and H.F. developed the concept and conceived experiments. N.K. supervised the project leading to this publication. K.O. wrote the manuscript. All authors contributed to the scientific discussions.
\renewcommand{\thetable}{S\arabic{table}}
\renewcommand{\thefigure}{S\arabic{figure}}
\setcounter{table}{0}
\setcounter{figure}{0}
\begin{figure*}[]
\centering
\includegraphics[width=13cm,clip]{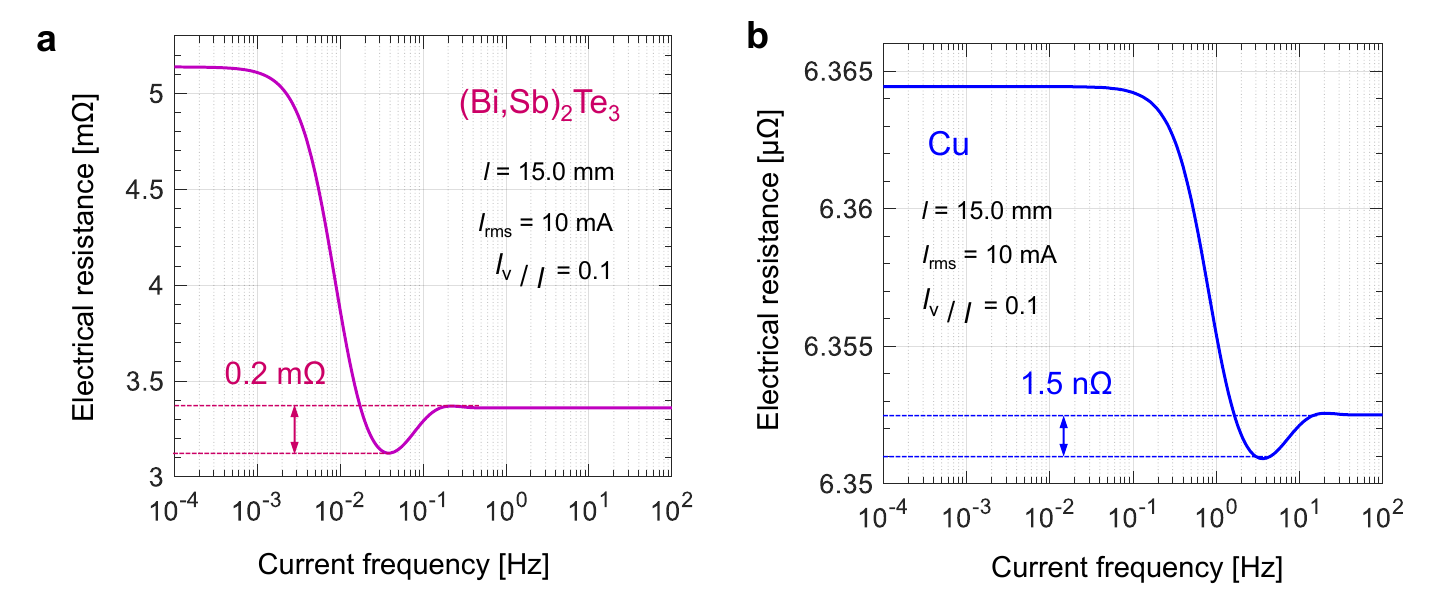}
\caption{{\bf Comparison between the electrical resistance change induced by the reverse heat flow of (Bi,Sb)$_{2}$Te$_{3}$ and Cu}. Calculation results based on the exact solution of the electrical resistance dependence of current frequency in (a) a thermoelectric material (Bi,Sb)$_{2}$Te$_{3}$ and (b) a conventional metallic Cu for the voltage probe position, $l_{\rm V}$, normalized by the sample length, $l$, $l_{\rm V}/l = 0.1$. Typical physical properties of (Bi,Sb)$_{2}$Te$_{3}$ (the thermal diffusivity $\alpha = \SI{1.25e-6}{m^{2}s^{-1}}$, Seebeck coefficient $S = \SI{165}{\micro VK^{-1}}$, and thermal conductivity $\kappa = \SI{1.45}{Wm^{-1}K^{-1}}$) and Cu ($\alpha = \SI{1.17e-4}{m^{2}s^{-1}}$, $S = \SI{6.5}{\micro VK^{-1}}$, and $\kappa = \SI{398}{Wm^{-1}K^{-1}}$) at room temperature were used in the calculation. The negative resistance change occurs in both materials, showing a resistance decrease of up to approximately $\SI{0.2}{m\Omega}$ at root-mean-square (rms) ac current $I_{\rm rms} = \SI{10}{mA}$ in (Bi,Sb)$_{2}$Te$_{3}$. However, Cu shows a resistance decrease of only $\SI{1.5}{n\Omega}$, which is $10^{-5}$ times that of (Bi,Sb)$_{2}$Te$_{3}$.}
\label{figS1}
\end{figure*}
\begin{figure*}[]
\centering
\includegraphics[width=10cm,clip]{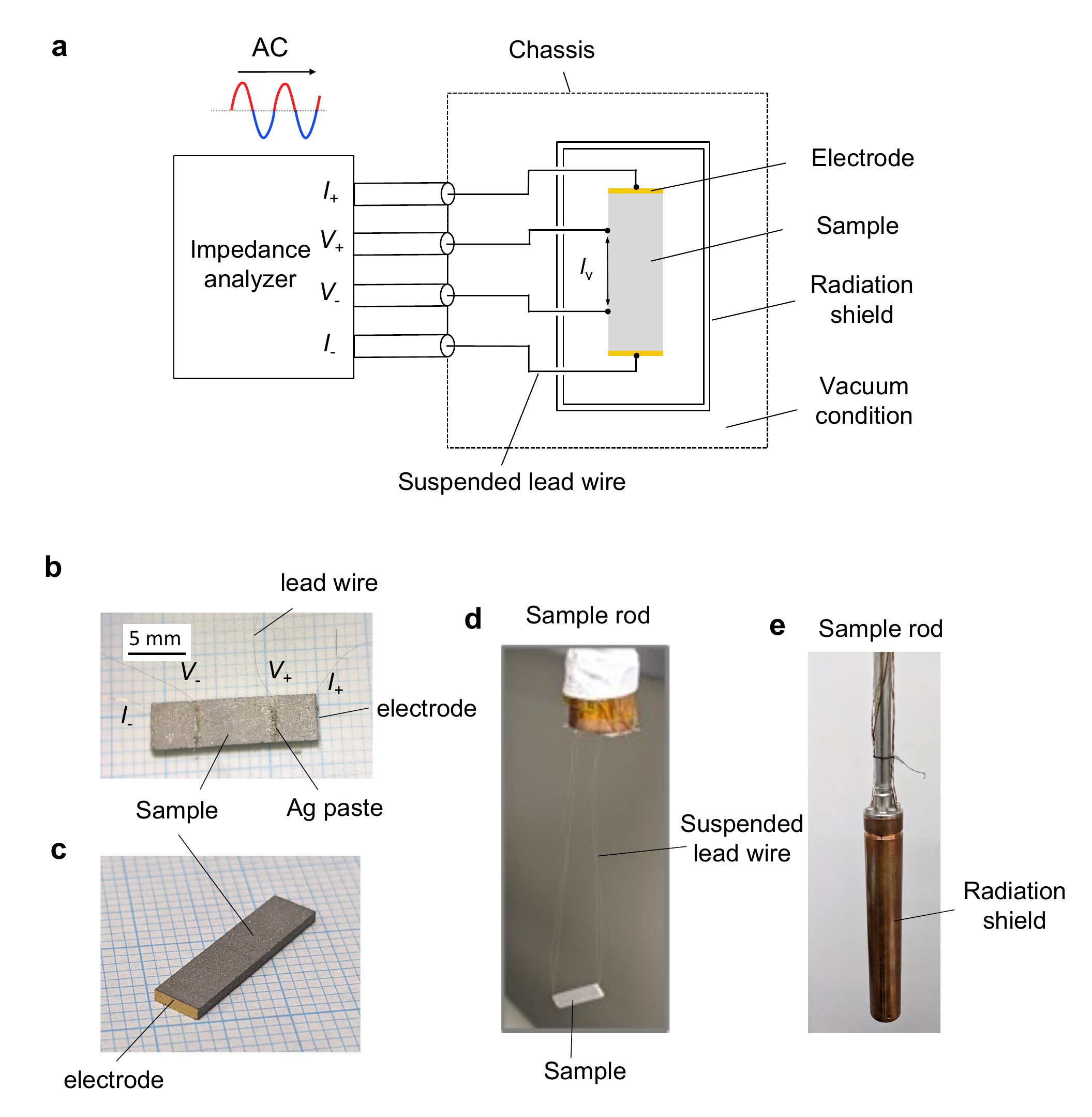}
\caption{{\bf Schematic of measurement setup.} (a) Measurement setup for the four-terminal pair configuration of the electrical resistivity measurement used in this study. (b), (c) Photographs of a polycrystalline p-type (Bi,Sb)$_{2}$Te$_{3}$ material. Photographs of (d) the setup of the suspended sample and (e) the radiation shield.
}
\label{figS2}
\end{figure*}
\begin{figure*}[]
\centering
\includegraphics[width=8cm,clip]{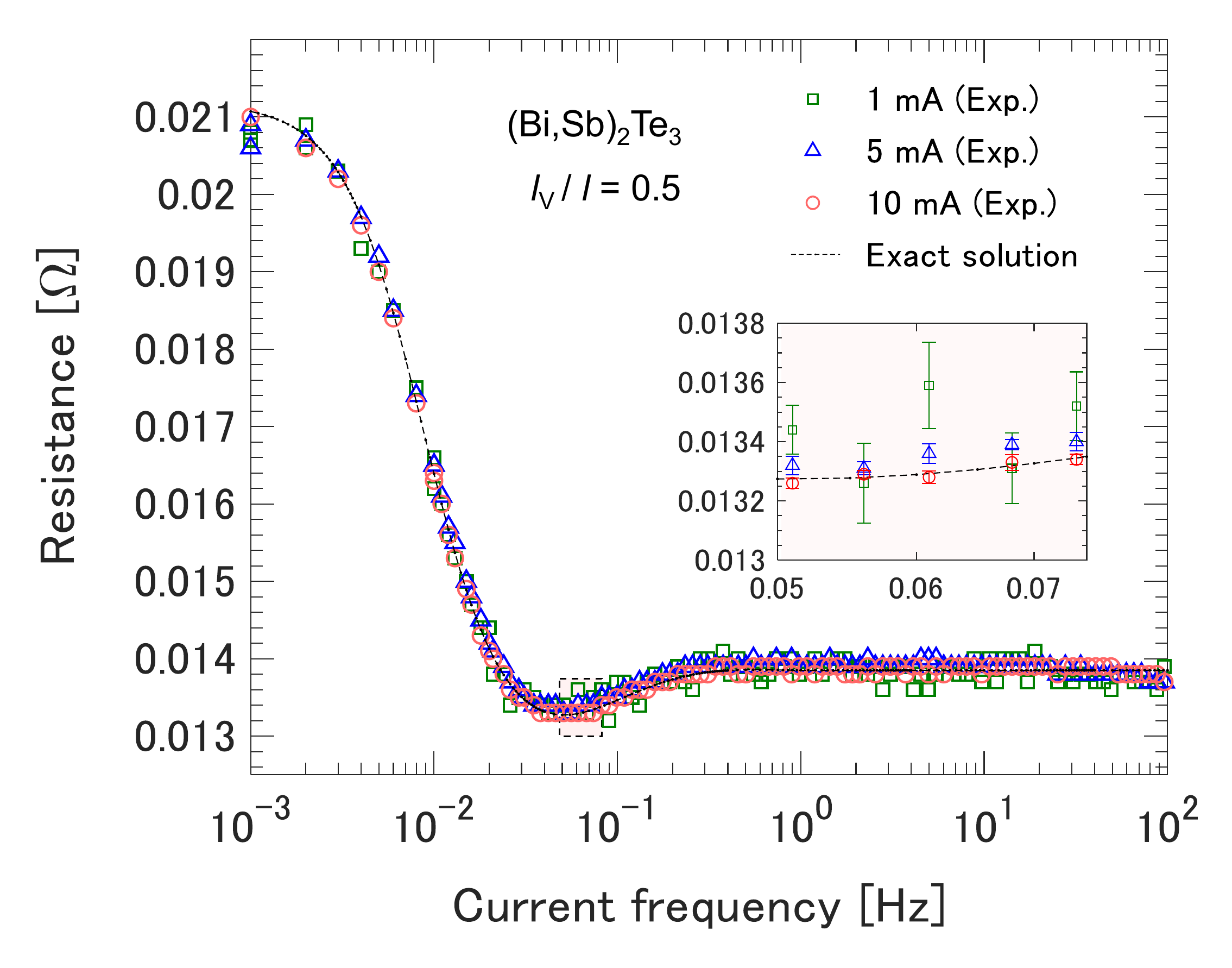}
\caption{{\bf Current amplitude dependence of resistance measurements.} The measurement results of the current frequency dependence of the resistance of (Bi,Sb)$_{2}$Te$_{3}$ with various current amplitude ($\SI{1}{mA}$, $\SI{5}{mA}$, and $\SI{10}{mA}$) at the voltage probe position normalized by the sample length $l_{\rm V}/l = 0.5$. The inset shows the enlarged view around $\SI{60}{mHz}$ with error bars. The error bars show the standard deviation of the mean for repeated measurements of the resistance.}
\label{figS3}
\end{figure*}
\begin{figure*}[]
\centering
\includegraphics[width=15cm,clip]{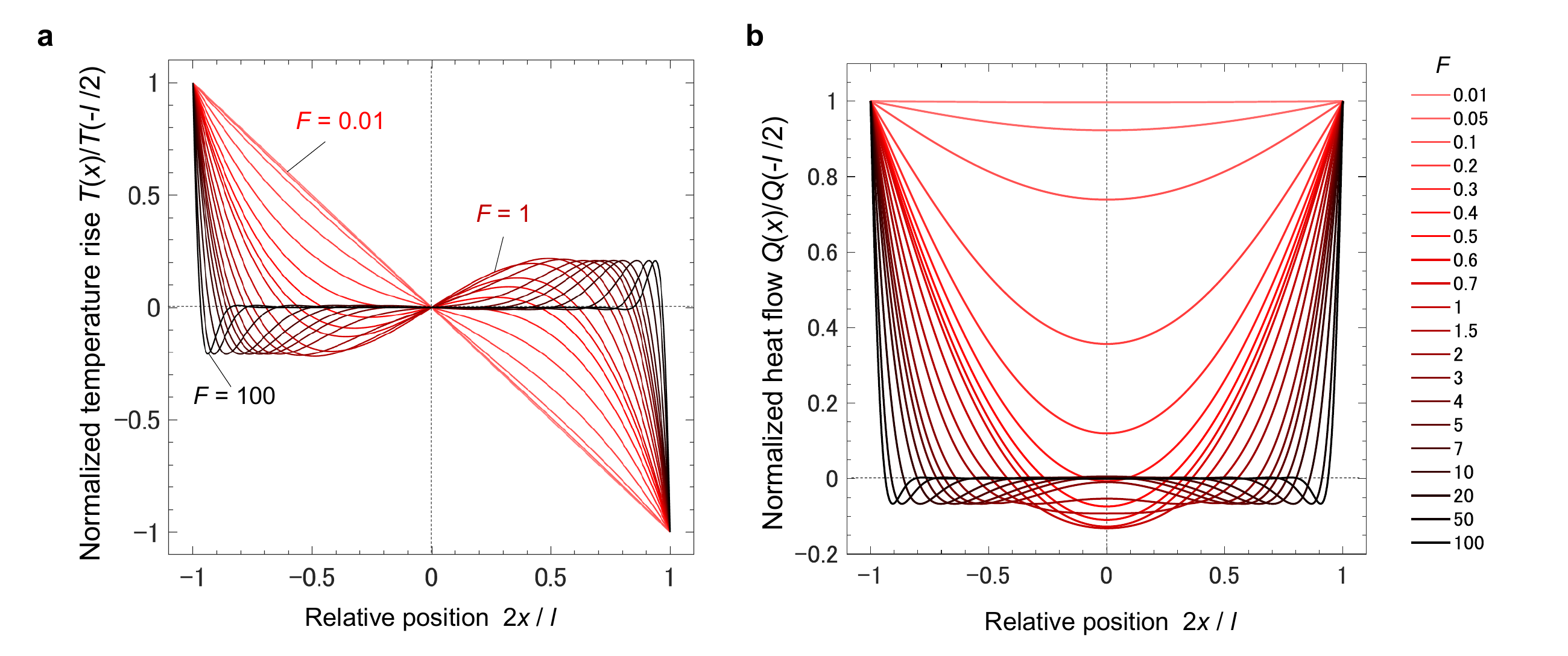}
\caption{{\bf Detailed simulation showing the relative position $2x/l$ dependence of the temperature $T(x)$ and heat flow $Q(x)$.} Calculation results based on the exact solution showing the dependence of relative position $2x/l$ on (a) temperature rise $T(x)$ and (b) heat flow $Q(x)$. The dimensionless parameter $F$ was varied from 0.01 to 100. The vertical axis is normalized by the value at the end of the sample, i.e., $x = l/2$.}
\label{figS4}
\end{figure*}
\begin{table*}
\caption{{\bf The parameters calculated by the exact solution in several materials (metals and semiconductors).}}
\begin{ruledtabular}
\begin{tabular}{cccccccc}
&Seebeck&Electrical&Thermal&&Negative&Negative&\\
Material&coefficient&resistivity&conductivity&$zT$ at&temperature&resistance&Frequency\\
 &($\SI{}{\micro VK^{-1}}$)&($\SI{}{\micro\ohm m}$)&($\SI{}{Wm^{-1}K^{-1}}$)&$\SI{300}{K}$&change (mK)&change ($\SI{}{\micro\ohm}$)&(Hz)\\ \hline
 (Bi,Sb)$_{2}$Te$_{3}$&165&11&1.5&0.53&25&200&0.046 \\
Ni-Cr alloy&21.7&0.68&16.7&0.012&0.25&0.42&0.14\\
 Cu-Ni alloy&-35&0.48&21.9&0.025&0.31&0.84&0.2\\
 Cu&6.5&0.017&398&0.0018&0.0035&0.0015&3\\
\end{tabular}
\end{ruledtabular}
\end{table*}

\nocite{*}

%\bibliography{apssamp}% 

\newpage

\begin{thebibliography}{99}
\expandafter\ifx\csname natexlab\endcsname\relax\def\natexlab#1{#1}\fi
\expandafter\ifx\csname bibnamefont\endcsname\relax
  \def\bibnamefont#1{#1}\fi
\expandafter\ifx\csname bibfnamefont\endcsname\relax
  \def\bibfnamefont#1{#1}\fi
\expandafter\ifx\csname citenamefont\endcsname\relax
  \def\citenamefont#1{#1}\fi
\expandafter\ifx\csname url\endcsname\relax
  \def\url#1{\texttt{#1}}\fi
\expandafter\ifx\csname urlprefix\endcsname\relax\def\urlprefix{URL }\fi
\providecommand{\bibinfo}[2]{#2}
\providecommand{\eprint}[2][]{\url{#2}}


\bibitem{Garimella1}
Garimella, S. V., Joshi, Y. K., Bar-Cohen, A., Mahajan, R., Toh, K. C., Carey, V. P., Baelmans, M., Lohan, J., Sammakia, B., Andros, F., Thermal challenges in next generation electronic systems. IEEE Trans. Components Packag. Technol. {\bf 31}, 801 (2008).

\bibitem{Ma2}
Ma, K., He, N., Liserre, M. \& Blaabjerg, F. Frequency-Domain Thermal Modeling and Characterization of Power Semiconductor Devices. IEEE Trans. Power Electron. {\bf 31}, 7183 (2016).

\bibitem{Ahmed3}
Ahmed, H. E., Salman, B. H., Kherbeet, A. Sh. \& Ahmed, M. I. Optimization of thermal design of heat sinks: A review. Int. J. Heat Mass Tranf. {\bf 118}, 129 (2018).

\bibitem{Shen4}
Shen, Y., Wang, H., Blaabjerg, F., Zhao, H., \& Long, T. Thermal Modeling and Design Optimization of PCB Vias and Pads. IEEE Trans. Power Electron. {\bf 35}, 882 (2020).

\bibitem{Li5}
Li, B., Wang, L., \& Casati, G. Thermal diode: Rectification of heat flux. Phys. Rev. Lett. {\bf 93}, 184301 (2004).

\bibitem{Ben6}
Ben-Abdallah, P., \& Biehs, S. A. Phase-change radiative thermal diode. Appl. Phys. Lett. {\bf 103}, 191907 (2013).

\bibitem{Martinez7}
Martínez-Pérez, M. J., Fornieri, A., \& Giazotto, F. Rectification of electronic heat current by a hybrid thermal diode. Nat. Nanotechnol. {\bf 10}, 303 (2015).

\bibitem{Li8}
Li, B., Wang, L., \& Casati, G. Negative differential thermal resistance and thermal transistor. Appl. Phys. Lett. {\bf 88}, 143501 (2006).

\bibitem{Sood9}
Sood, A., Xiong, F., Chen, S., Wang, H., Selli, D., Zhang, J., McClellan, C. J., Sun, J. Donadio, D., Cui, Y., Pop,E., \& Goodson, K. E. An electrochemical thermal transistor. Nat. Commun. {\bf 9}, 4510 (2018).

\bibitem{Wang10}
Wang, L., \& Li, B. Thermal Logic Gates: Computation with phonons. Phys. Rev. Lett. {\bf 99}, 177208 (2007).

\bibitem{Wang11}
Wang, L., \& Li, B. Thermal memory: A storage of phononic information. Phys. Rev. Lett. {\bf 101}, 267203 (2008).

\bibitem{Kubytskyi12}
Kubytskyi, V., Biehs, S.-A., \& Ben-Abdallah, P. Radiative bistability and thermal memory. Phys. Rev. Lett. {\bf 113}, 074301 (2014).

\bibitem{Guarcello13}
Guarcello, C., Solinas, P., Braggio, A., Ventra, M. D., \& Giazotto, F. Josephson thermal memory. Phys. Rev. Applied {\bf 9}, 014021 (2018).

\bibitem{Li14}
Li, N., Re, J., Wang, L., Zhang, G., Hanggi, P., \& Li, B. Colloquium: Phononics: Manipulating heat flow with electronic analogs and beyond. Rev. Mod. Phys. {\bf 84}, 1045 (2012).

\bibitem{Wehmeyer15}
Wehmeyer, G., Yabuki, T., Monachon, C., Wu, J., \& Dames, C. Thermal diodes, regulators, and switches: Physical mechanisms and potential applications. Appl. Phys. Rev. {\bf 4}, 041304 (2017).

\bibitem{Ye16} 
Ye, H., Leung, S. Y. Y., Wong, C. K. Y., Lin, K., Chen, X., Fan, J., Kjelstrup, S., Fan, X., \& Zhang, G. Thermal inductance in GaN devices. IEEE Electron Device Lett. {\bf 37}, 1473 (2016).

\bibitem{Bosworth17}
Bosworth, R. C. L. The thermal Ohm, Farad and Henry. Philos. Mag. {\bf 37}, 803 (1946). 

\bibitem{Weedy18}
Weedy, B. M. The analogy between thermal and electrical quantities. Elector. Pow. Syst. Res. {\bf 15}, 197 (1988). 

\bibitem{Lienhard19}
Lienhard, J. H. V, \& Lienhard, J. H. IV, A Heat Transfer Textbook: Fifth Edition (Dover Books on Engineering), (2019).

\bibitem{Piggott20}
Piggott, A. J., \& Allen, J. S. Peltier Supercooling with Isosceles Current Pulses: A Response Surface Perspective. ECS J. Solid State Sci. Technol., {\bf 6}, N3045 (2017). 

\bibitem{Bossen21}
Bossen, O. \& Schilling, A. LC-circuit calorimetry. Rev. Sci. Instrum. {\bf 82}, 094901 (2011). 

\bibitem{Bosworth22}
Bosworth, R. C. L. Thermal inductance. Nature {\bf 158}, 309 (1946).

\bibitem{Bosworth23}
Bosworth, R. C. L. Thermal mutual inductance. Nature {\bf 161}, 166 (1948).

\bibitem{Schilling24}
Schilling, A., Zhang, X. \& Bossen, O. Heat flowing from cold to hot without external intervention by using a “thermal inductor”. Sci. Adv. {\bf 5}, eaat9553 (2019).

\bibitem{Iwasaki25}
Iwasaki, H., \& Hori, H. Thermoelectric property measurement by the improved Harman method. in International Conference on Thermoelectrics, ICT Proceedings, pp.501 (2005).

\bibitem{Downey26}
Downey, A. D., Hogan, T. P., \& Cook, B., Characterization of thermoelectric elements and devices by impedance spectroscopy. Rev. Sci. Instrum. {\bf 78}, 093904 (2007).

\bibitem{Garcia27}
Garcia-Canadas, J., \& Min, G., Low Frequency Impedance Spectroscopy Analysis of Thermoelectric Modules. J. Electron. Mater. {\bf 43}, 2411 (2014).

\bibitem{Beltrani28}
Beltrán-Pitarch, B., Prado-Gonjal, J., Powel, A. V., \& García-Cañadas, J. Experimental conditions required for accurate measurements of electrical resistivity, thermal conductivity, and dimensionless figure of merit (ZT) using Harman and impedance spectroscopy methods. J. Appl. Phys. {\bf 125}, 025111 (2019).

\bibitem{Shinozaki29}
Shinozaki, R., Hirabayashi, S., \& Hasegawa, Y. Dimensionless figure of merit of constantan estimated using impedance spectroscopy. Appl. Phys. Express. {\bf 13}, 106501 (2020).

\bibitem{Amagai31}
Amagai, Y., Shimazaki, T., Okawa, K., Kawae, T., Fujiki, H., \& Kaneko, N.-H. High-accuracy compensation of radiative heat loss in Thomson coefficient measurement. Appl. Phys. Lett. {\bf 117}, 063903 (2020).

\bibitem{Kirby30}
Kirby, C. G. M., \& Laubitz, M. J. The error due to the Peltier effect in direct-current measurements of resistance. Metrologia {\bf 9}, 103 (1973). 

\end{thebibliography}

\end{document}